\documentclass[twocolumn,prd,superscriptaddress]{revtex4}
\usepackage[pdftex]{graphicx}
\usepackage{amsmath, amsfonts, amssymb}
\usepackage{slashed}
\usepackage[T1]{fontenc}
\usepackage{txfonts}
\usepackage{xcolor}
\newcommand{\be}{\begin{equation}}
\newcommand{\ee}{\end{equation}}
\newcommand{\B}[1]{\mathbf{#1}}
\newcommand{\BS}[1]{\boldsymbol{#1}}
\newcommand{\C}[1]{\mathcal{#1}}
\usepackage{multirow}
\graphicspath{{figures/}}

\newcommand{\bhat}[1]{\hat{\mathbf{#1}}}

\begin{document}
\title{Fundamental Relations between Measurement, Radiation and Decoherence in Gravitational Wave Laser Interferometer Detectors}
\author{Belinda Pang}
\affiliation{Theoretical Astrophysics and Walter Burke Institute for Theoretical Physics, M/C 350-17, California Institute of Technology, Pasadena, California 91125}
\author{Yanbei Chen}
\affiliation{Theoretical Astrophysics and Walter Burke Institute for Theoretical Physics, M/C 350-17, California Institute of Technology, Pasadena, California 91125}

\begin{abstract}
As laser interferometer gravitational wave (GW) detectors become quantum noise dominated, understanding the fundamental limit on measurement sensitivity imposed by quantum uncertainty is crucial to guide the search for further noise reduction.  Recent efforts have included applying ideas from quantum information theory to GW detection -- specifically the quantum Cramer Rao bound, which is a minimum bound on error in parameter estimation using a quantum state and is determined by the state's quantum Fisher information (QFI) with respect to the parameter~\cite{helstrom1969}. Identifying the QFI requires knowing the interaction between the quantum measurement device and the signal, which was rigorously derived for GW interferometer detectors in Ref.~\cite{pang2018}. In this paper, we calculate the QFI and fundamental quantum limit (FQL) for GW detection, and furthermore derive explicit reciprocity relations involving the QFI which summarize information exchange between the detector and a surrounding weak quantum GW field. Specifically, we show that the GW power radiation by the detector's quantum fluctuations are proportional to the QFI, and therefore inversely proportional to its FQL. Similarly, the detector's decoherence rate in a white noise GW bath can be explicitly related to the QFI/FQL. These relations are fundamental and appear generalizable to a broader class of quantum measurement systems.
\end{abstract}
\maketitle

\section{Introduction}

The detection of gravitational waves (GWs) by the Laser Interferometer Gravitational-wave Observatory (LIGO) is a remarkable achievement which confirmed a key prediction of general relativity (GR). Together, detections by the LIGO-VIRGO network made the first direct observation of binary black holes, provided the first direct link between binary neutron star inspirals and short $\gamma$-ray bursts, probed the nature of GW polarizations and allowed for additional tests of GR \cite{GW150914, GW170814,GW170817} .  These  advancements mark only the beginning of multi-messenger astronomy with gravitational waves and investigations into the properties of spacetime in the strong-field dynamical regime~\cite{barack2018black}.   The community has conceived major upgrades that improve sensitivities across the entire detection band, e.g., LIGO Voyager,  the Einstein Telescope~\cite{punturo2010einstein}, the Cosmic Explorer~\cite{abbott2017exploring}, as well as those that target mainly higher~\cite{miao2018towards} and lower frequencies~\cite{yu2018prospects}.

Significant improvements in detector sensitivity have resulted from reducing classical noise sources, such as ground vibration (which couples into the gravitational-wave readout via many linear and nonlinear channels) and thermal fluctuations, with continuing effort in this direction. However, as these noise sources become suppressed and detectors approach the quantum-limited regime, further improvements in detection sensitivity that are required to access farther reaches of the universe will necessitate the manipulation of noise arising from quantum uncertainty, specifically the quantum fluctuations of light and test masses \cite{adhikari2014gravitational,danilishin2012quantum}. In contrast to classical noise sources, these quantum fluctuations cannot be eliminated even in principle, but they can be manipulated to reduce their effect on the overall signal-to-noise ratio.

Since LIGO's inception, the gravitational wave (GW) community's understanding of quantum noise for interferometer detectors has seen steady progression. The first identification of a quantum limit was from the direct application of the Heisenberg uncertainty principal to the continuous monitoring of the position of the test masses~\cite{caves1980measurement,braginskyQM}. It was later realized that this so-called standard quantum limit (SQL) can be beaten in different ways, including: (i) modifying the input quantum state of the optical field and/or the way the out-going field is read out~\cite{kimble2001conversion}, and/or (ii) modifying the optomechanical dynamics of the interferometer~\cite{braginsky1999low,buonanno2001quantum,buonanno2002signal}.  Intuitively, we can surpass the SQL either via establishing/utilizing quantum correlations between the sensing and back-action noise, or by modifying the dynamics so that the test masses are no longer free masses~\cite{danilishin2012quantum}.  

With the SQL no longer a strict limit, it is important to find   {a more relevant} fundamental limit for GW detection.  Braginsky, Gorodetsky, Khalili and Thorne (BGKT) proposed the {\it Energetic Quantum Limit} (EQL)~\cite{braginskyEQL}, an upper limit for the detector's {\it signal-to-noise ratio} in terms of the spectrum of quantum fluctuations of the in-cavity optical energy (see also Sec.~9.2 of \cite{braginsky1995quantum}),

Tsang, Wiseman and Caves (TWC) later obtained the Quantum Cramer-Rao Bound (QCRB) for waveform estimation, showing that the EQL is indeed a {\it Fundamental Quantum Limit} (FQL) for linear measuring devices at steady state~\cite{tsang2011}.  
Miao {\it et al.}~\cite{HM17,branford2018fundamental}  considered a broad class of gravitational-wave detector dynamics, showing explicitly that: (i) the signal-to-noise ratio of these detectors are indeed bound by the FQL, (ii) the FQL can be increased by modifying the input quantum state of the optical field and by modifying the system's optomechanical dynamics, and (iii) one can always reach within a factor of 2 (in power) of the FQL, if a (usually frequency dependent) homodyne detection can be performed on the out-going optical field.  

In the work leading to the FQL for GW detectors, it was most convenient to treat the GW as directly coupled to the amplitude quadrature of the optical field inside the arm cavities --- instead of as applying a tidal force on the test masses.   In a previous work, we derived the interactions between a laser interferometer and GWs from first principles \cite{pang2018}, providing a solid foundation for such a treatment.  It is the first aim of this paper to present the rigorous FQL for gravitational-wave detectors. It has been suggested, by Levin,  that the FQL is related to the power at which the GW detector, driven {\it only by quantum fluctuations}, radiate GWs~\footnote{Y.\ Levin, private communications}.  We shall confirm this {\it reciprocity relation} by putting it into an explicit and quantitative form, as the second aim of our paper.

Another effect that arises naturally as we treat GW as quantum is that the device will suffer from {\it gravitational decoherence}. We will show that, in fact, like (quantum-driven) radiation power, the decoherence rate of the detector is also directly related to the FQL.  While the reciprocity between radiation and detection has a classical analogy to radio antenna \cite{JC24},  the relationship between the fundamental bound on measurement sensitivity and the quantum decoherence of the measurement device has not, to the best of our knowledge, been quantified before. 

The {\it reciprocity} relations between FQL, energy radiated and decoherence lead to a conceptual issue: because the radiation emitted by the quantum-driven detector is quantum, we are treating GWs as quantum --- even though we were applying the QCRB, which treats GW as classical.   We would like to argue that  the radiation and decoherence effects here shows the limitation of the QCRB formalism: ultimately the signal we are detecting is quantum, and it will be an internal consistency that the radiation by the detector has a power spectrum much less compared with vacuum fluctuations of the signal itself, and that decoherence due to coupling to the signal is small compared with other decoherence of the device, especially the one induced by quantum measurement itself. 

This paper is organized as follows.  In Sec.~\ref{sec:review}, we provide a more detailed review of sensitivity limits of gravitational-wave detectors.  In Sec.~\ref{sec:QCRB} we briefly review the interaction between GW and the laser interferometer derived in \cite{pang2018} and use the result to obtain the QCRB. Finally, in Sec.~\ref{sec:Reciprocity} we discuss the reciprocity relations between FQL, radiation, and decoherence.

\section{Sensitivity Limits for Gravitational-wave Detection}
\label{sec:review}

In this section, we provide a (historical) overview of quantum limits for GW detectors, as well as more recent  motivations for considering the fundamental quantum limit. 

\subsection{Standard Quantum Limit}

The understanding of quantum noise in laser interferometer gravitational-wave detectors has evolved since Braginsky's formulation of the Standard Quantum Limit (SQL) for high precision measurements \cite{braginskyQM}. 

In the simplest picture (see, e.g., Ref.~\cite{kimble2001conversion}) of quantum noise in a gravitational-wave detector (e.g., Fabry-Perot Michelson interferometer),  the input laser beam is in a displaced coherent state, and drives a single optical mode of each of the arm cavities.  Differential quantum fluctuations in the cavity modes, which are responsible for the detector's quantum noise, are injected into the interferometer through its antisymmetric port.  The detector suffers from two types of noise:  radiation pressure noise, due to motions of mirrors driven by amplitude fluctuations of light (which act as a {\it ponderomotive force}), and shot noise, due to discreteness of photons when they arrive at the photodetectors.  If the outgoing field is measured directly along its phase quadrature (which carries the gravitational wave signal), then the measured shot noise will be due to phase fluctuations.

Since the signal to noise ratio for a Poisson process goes as $1/\sqrt{N}$ (with $N$ the number of photons), to decrease shot noise one would have to increase laser power. However, doing so would increase the radiation pressure noise. Thus, we see that there is a tradeoff between the two types of quantum noise which implies that we cannot make both arbitrarily small. This is in fact a manifestation of the Heisenberg uncertainty principle, since the radiation pressure and shot noise are associated with conjugate quadratures of the cavity mode (phase and amplitude, represented by $\hat{\alpha}_1$ and $\hat{\alpha}_2$) which satisfy the canonical commutation relation $[\hat{\alpha}_1,\hat{\alpha}_2]=i\hbar$. 

In the frequency domain, radiation pressure noise dominate at low frequencies while shot noise dominates at high frequencies. Heuristically speaking, this frequency dependence can be attributed to the fact that radiation pressure follows the frequency response of the test mass, which goes as $1/\Omega^2$, while the shot noise is frequency independent.  This tradeoff between radiation pressure and shot noise as one adjusts the input power results in a set of curves, the locus of whose minima results in the standard quantum limit.  For a Fabry-Perot Michelson interferometer withe arm length $L$ and four mirrors with mass $M$, the free-mass SQL for the gravitational-wave strain $h$, in terms of noise spectrum at angular frequency $\Omega$,  is given by  
\begin{equation}
\label{sqlfm}
S_h^{\rm SQL}={\frac{8\hbar}{M\Omega^2 L^2}}
\end{equation}   

Although the SQL depends only on fundamental quantities, it is not, in fact, a fundamental sensitivity limit.  After Caves showed that detector noise can be suppressed by injecting squeezed vacuum from the dark port~\cite{caves1981}, Uhruh showed that the appropriate squeezed vacuum can allow the detector to surpass the SQL at certain frequencies~\cite{unruh1983}.  Kimble {\it et al.} further proposed that frequency-dependent squeezed vacuum, produced by filtering frequency-independent squeezed vacuum through a detuned cavity, allows the detector to surpass the SQL globally~\cite{kimble2001conversion}.  It was later shown that frequency-dependent squeezed vacuum can also be produced when injecting entangled squeezed vacuum into the interferometer's dark port, performing separate homodyne detections after they return, and making the appropriate combination of measurement results~\cite{YM17}. 

On modification of the readout schemes, Vyatchanin and Matsko showed that for a signal with known shape and arrival time, an appropriate choice of a time-dependent readout quadrature allows us to eliminate back-action noise~\cite{vyatchanin1996}.  Such {\it back-action evasion} was later shown to be possible at all frequencies, e.g., by Kimble {\it et al.}, if a frequency-dependent homodyne detection is performed~\cite{kimble2001conversion}. 

Yet another way to circumvent the SQL was to modify the dynamics of the test-mass mirrors in the interferometer~\cite{braginsky1999low,buonanno2001quantum,buonanno2002signal}. This was shown to be possible in Fabry-Perot Michelson interferometers with signal recycling.  In these detectors, differential optical powers in the arms depend on the position of the mirrors, creating an ``optical spring'', and amplifying the mirror's response to  gravitational waves near the spring resonant frequency. The detector can surpasses the free-mass SQL~\eqref{sqlfm}, because the SQL for oscillators (in terms of noise spectrum) is much lower.

\subsection{Mizuno Bound, Energetic Quantum Limit, and White-Light Cavities}
With the SQL no longer being a fundamental limit, we need to search for a new, and more relevant limit. Since back-action noise is shown to be avoidable, one can focus on shot noise. 

One such bound, for interferometers with infinite masses (i.e., ignoring radiation pressure effects), was obtained by Mizuno, when considering signal recycling and resonant side-band extraction interferometers. It was shown that such interferometers can only {\it reshape} the noise curve:  the peak sensitivity over certain frequencies can be increased  at the expense of detection bandwidth \cite{meers1988recycling, mizuno1993resonant}. Mizuno pointed that the tradeoff can by summarized by the statement that the area under the noise curve must be conserved.
\begin{equation}\label{mizunoLimit}
\int \frac{1}{S_h} \frac{d\Omega}{2\pi} \le N \omega_0^2
\end{equation} 
where $\omega_0$ the resonant frequency and $N$ is the mean number of intracavity photons.

{A more general EQL was obtained by BGKT~\cite{braginskyEQL} when considering the quantum uncertainty principle between energy and phase (the phase being the carrier of the GW signal). The EQL states that
\begin{equation}\label{EQL}
\frac{1}{S_h} \le \frac{S_{\mathcal{E}}}{\hbar^2}
\end{equation}
where $S_{\mathcal{E}}$ is the noise spectrum for energy, which evaluates to the Mizuno bound when the cavity's optical mode is in a coherent state. In this case, integrating the right-hand side of Eq.~\eqref{EQL} over frequency gives the result
\begin{equation}\label{coherentEQL}
\int \frac{S_{\mathcal{E}}}{\hbar^2} \frac{d\Omega}{2\pi} =\frac{\langle\Delta \mathcal{E}^2\rangle}{\hbar^2} =\frac{\mathcal{E^2}}{\hbar^2 N}
\end{equation} 
where $\mathcal{E}$ is the mean intracavity energy (note that the RHS of Eq.~\eqref{coherentEQL} is equal to that of Eq.~\eqref{mizunoLimit}). The EQL is based on a derivation for the minimum detectable force in general linear quantum measurements, in which the authors concluded that the maximum signal to noise ratio is determined by fluctuations of the interaction energy~\cite{braginskyQM9}.}

Interest in the gravitational-wave community on fundamental limits were re-ignited due to renewed research efforts into {\it white-light cavities} (WLCs), so called because they eliminate the frequency-dependent optical phase delay, which would enable higher broadband sensitivity.  Wicht {\it el al.}~\cite{wichtWLC1997, wichtWLC2007} were the first to propose placing an atomic gain medium with {\it anomalous dispersion} into the optical cavity to cancel the frequency dependent propagation phase of light, producing a simultaneous improvement of peak sensitivity and bandwidth, which seemingly violates the naive application of EQL. More recently, Pati, Yum {\it et al.}~\cite{PatiWLC2007, yumWLC2013} have proposed different types of active media to achieve this same effect.  Zhou {\it et al.}  studied the application of active medium to LIGO-type interferometers.~\cite{zhou2015quantum} 

Independently, it was proposed to use grating systems to realize anomalous dispersion, but this was shown not to work~\cite{wise2005phase}.  Discussions in Ref.~\cite{wise2005phase} indicated that anomalous dispersion generally requires active amplification processes which was absent in the grating system. However, it has long been known that amplification processes in the quantum regime will bring additional noise~\cite{caves1982quantum}. In particular, as noted by Kuzmich {\it et al..}~\cite{kuzmich2001}, the use of an active medium necessarily introduces quantum noise associated with pumping.  Furthermore, Ma {\it et al.}~\cite{ma2015} showed that, for the configuration proposed in Ref.~\cite{yumWLC2013}, in the parameter regime where the active medium is stable, there is no net enhancement in shot-noise limited sensitivity --- and it was not clear at the time whether the system can operate in the unstable regime.   The existence of additional noise and possible instability makes it unclear whether WLCs (which are conceptually grounded in anomalous dispersion) would be viable techniques for improving sensitivity.

Later, Miao {\it et al.}~\cite{miaoWLC2015} proposed using an unstable optomechanical filter to implement the propagation phase cancellation, and in this case the system was shown to be controllable without sacrificing sensitivity.  Zhou {\it et al.} proposed further optomechanical realizations of negative dispersion~\cite{zhou2018optomechanical}.

The above discussions strongly motivates a better understanding not only of the origin of the EQL, but also of when it is achievable. 

\subsection{Quantum Cramer-Rao Bound and Fundamental Limit for Waveform Detection}

A formal way to deduce the fundamental quantum limit for gravitational-wave detection comes  from the field of quantum parameter estimation in the form of the quantum Cramer-Rao bound (QCRB). In analogy with the classical Cramer-Rao bound, the QCRB formally states that the minimum variance of the optimal estimator for a classical quantity parametrizing a quantum probe is the inverse of the probe's quantum Fisher information (QFI), maximized over all possible positive operator valued measures (POVMs). A more detailed derivation follows.

\subsubsection{QCRB for a scalar parameter}

For a scalar parameter $\lambda$ and a density matrix $\rho(\lambda)$ that depends on $\lambda$, we construct the {\it Quantum Fisher Information}
\be\label{QFIDef}
F_{\lambda\lambda} = {\rm Tr} \left[L_\lambda^\dagger L_\lambda \rho(\lambda)\right]
\ee
where $L_\lambda$ is the logarithmic derivative operator defined such that
\be
\frac{L_\lambda \rho(\lambda) + \rho(\lambda)  L_\lambda^\dagger}{2} = \frac{\partial \rho(\lambda)}{\partial \lambda}
\ee
Then for any unbiased estimator $X$ for $\lambda$ which satisfies
$
\mathrm{tr} [X\rho(\lambda)] =\lambda
$,
the estimation error satisfies  the Quantum Cramer Rao Bound (QCRB) of 
\be
\label{eqQCRB}
\sigma_{\lambda\lambda}^2 \equiv \mathrm{tr} \left[(X-\lambda)^2\rho(\lambda)\right]  \geq \frac{1}{F_{\lambda\lambda}}
\ee

An important special case is when $\lambda$ is the amplitude of a unitary transformation on an initial pure state $\rho_0$ such that 
{\begin{equation}
\rho(\lambda) = e^{-i\lambda\hat{G}/\hbar} \rho_0 e^{i\lambda\hat{G}/\hbar}
\end{equation}
where $G$ is a Hermitian operator. One can easily obtain 
\begin{equation}
L_\lambda = -2i  (\hat G-c)
\end{equation}
where $c$ can be any real number.  It can then be shown using Eq.~\eqref{QFIDef} that
\be
F_{\lambda\lambda} = \frac{4}{\hbar^2} \mathrm{tr}\left[ (\hat G-c)^2\rho(\lambda)\right] \equiv \frac{4}{\hbar^2} \left\langle (\hat G-c)^2\right\rangle\,.
\ee
Note that $\langle \cdot \rangle$ has been used to denote the expectation value taken with respect to $\rho(\lambda)$, which is the density matrix at the {\it true value} of $\lambda$. The parameter $c$ in $F_{\lambda\lambda}$ is arbitrary, which means that Eq.~\eqref{eqQCRB} holds true for all possible values.  The most stringent bound is obtained when $F_{\lambda\lambda}$ is at its minimum for
$
c=\mathrm{tr}\left[\hat G\rho(\lambda)\right] =\left\langle \hat G\right\rangle
$. {In this way, we simply define
\begin{equation}
F_{\lambda\lambda} =\frac{4}{\hbar^2 }\langle \Delta \hat G^2\rangle 
\end{equation} 
with $\Delta \hat G \equiv \hat G-\left\langle\hat G\right \rangle$. 
To summarize in words, for this simple case where $\rho_\lambda$ is obtained from a unitary transformation on an pure initial state $\rho_0$, the QFI is given by the variance of the generator $\hat{G}$ (so-called because it generates a translation of the quantum state proportional to $\lambda$) with respect to $\rho(\lambda) $. }

\subsubsection{QCRB for a list of parameters}

This framework can be extended to multiple parameters $\lambda_j$,  by identifying the operator $L_j$ corresponding to each parameter $\lambda_j$ in the vector $\BS{\lambda}$:
\begin{equation}
\partial_j \rho \equiv \frac{\partial\rho}{\partial \lambda_j} = \frac{L_j\rho+\rho L_j^\dagger}{2}\,.
\end{equation}
In this case, the {\it Quantum Fisher Information Matrix} (QFIM) is given by 
\begin{equation}
F_{jk} = {\rm Tr}\left[L_j^\dagger  L_k \rho(\lambda)\right]
\end{equation}
whose inverse bounds the covariance matrix $\sigma_{jk}$ of  unbiased estimators $X_j$ for $\lambda_j$, or
\begin{equation}
\sigma_{jk} \succ F_{ij}^{-1}\,,
\end{equation}
in the sense that the quantity $\sigma_{jk} - F_{ij}^{-1}$ is a positive definite quadratic form. In particular, the eigenvalues of $\sigma_{jk}$ and those of $F_{ij}^{-1}$ can be arranged in such a way that each of the former is greater than the corresponding eigenvalue of $F_{ij}^{-1}$. In the particular case of a unitary transformation where
\begin{equation}
\rho(\lambda) = e^{-i \sum\lambda_j \hat G_j /\hbar}\rho_0 e^{i \sum\lambda_j \hat G_j /\hbar}
\end{equation}
we have
\begin{equation}
F_{jk} = \frac{4}{\hbar^2 }\left\langle \Delta \hat G_j \,\Delta\hat G_k\right\rangle 
\end{equation} 

\subsubsection{QCRB for waveform estimation}

Tsang {\it et al.}~\cite{tsang2011} applied quantum multi-parameter estimation to obtain the QCRB for a continuous waveform $x(t)$.  {One key insight in this work is the {\it principle of deferred measurement}, which states that a series of measurements performed {\it during} the time evolution of a system can always be performed by a set of commuting operators on its {\it final state}. In this way, the QCRB for measuring $x(t)$ can be obtained if we know the dependence of the final density matrix $\rho_{\rm fin}$ on $x(t)$.  More specifically, let us write
\begin{equation}
H(t) = H_0(t) + G(t) x(t)
\end{equation}
where $H_0(t)$ and $G(t)$ are Schr\"odinger operators.  Suppose the evolution is from $0$ to $T$, we can write
\begin{equation}
\rho_{\rm fin} = \rho[T|x(t)] = U[x(t)] \rho (0) U^\dagger[x(t)]
\end{equation}
where $U[x(t)]$ is the evolution operator of the system from its initial to final state as a functional of $x(t)$.  Making a variation in $x(t)$, we can write 
\begin{equation}
\frac{\delta\rho_{\rm fin}}{\delta x(t_0)} =\frac{\delta U[x(t)]}{\delta x(t_0) }\rho(0) U^\dagger[x(t)] + U[x(t)] \rho(0) \frac{\delta U^\dagger[x(t)]}{\delta x(t_0) }   
\end{equation}
Writing
\begin{equation}
U[x(t)] = U[T,t_0+\Delta t] U[t_0+\Delta t,t_0-\Delta t] U[t_0-\Delta t ,0]
\end{equation} 
We first take variation and let $\Delta t\rightarrow 0$, and obtain
\begin{equation}
\frac{\delta U[x(t)] }{\delta x( t_0)}=-\frac{i}{\hbar}  U[T,t_0] G(t_0) U[t_0 ,0]
\end{equation}
Let us define 
\begin{equation}
G_H(t_0) \equiv U[T,t_0] G(t_0) U^\dagger[T,t_0]
\end{equation}
namely the {\it Heisenberg Operator} that corresponds to the result of evolution of  the Schr\"odinger operator $G(t_0)$, up till the end time $T$.  In terms of $G_H(t_0)$, we have
\begin{equation}
\frac{\delta\rho_{\rm fin}}{\delta x(t_0)} = -\frac{i}{\hbar}\left[ 
G_H(t_0) \rho_{\rm fin}  - \rho_{\rm fin} G_H(t_0)\right]   
\end{equation}
In this way, viewing $x(t)$ as the continuum limit of a vector, they obtain a continuous QFIM of
\be\label{QFIM}
F(t,t')=\frac{4}{\hbar^2}\langle \Delta \hat{G}_H(t)\Delta\hat{G}_H(t')\rangle_{0,\rm sym}
\ee
where the $H$ superscript indicates the Heisenberg picture operator and  the `${\rm sym}$' subscript denotes symmetrization. This general result for the QFIM of a continuous waveform only requires that the signal being detected appears explicitly in the Hamiltonian. The covariance $\Sigma$ of the estimator must then satisfy 
\begin{equation}
\Sigma(t, t')\succ F^{-1}(t, t').
\end{equation} }

In this paper, using the Hamiltonian developed in \cite{pang2018} which describes the interaction between laser interferometer gravitational-wave detectors (GW detectors for short) and gravitational waves and the above framework for waveform estimation developed in Ref.~\cite{tsang2011}, we obtain the QCRB for these detectors. We note that the QCRB has previously been considered for such detectors by Downes {\it et al. }~\cite{downes2017} but their results were derived for free electromagnetic field. In contrast, our derivation of the interaction Hamiltonian takes into account the interferometric configuration of the detector and the confinement of the cavity mode within a Fabry-Perot cavity. Our results provide theoretical confirmation for Miao {et al.'s}~\cite{HM17} work which showed that the QCRB for GW detectors can always be obtained up to a factor of $\sqrt{2}$, and which also discussed the relation of the QCRB with the noise reducing schemes mentioned in Sec.~\ref{sec:review}.

\section{QCRB for a Laser Interferometer Gravitational Wave Detector}
\label{sec:QCRB}

{Let us consider a} Michelson interferometer with Fabry-Perot cavity arms, which in the Newtonian gauge point of view detects gravitational waves on the principle that an incoming wave will cause a differential change in the lengths of the two cavities.   In this section, we shall first review our previously obtained Hamiltonian, and then write down the corresponding QCRB.

\subsection{Interaction Hamiltonian}

Since the QCRB only involves the observable that couples to the parameters we would like to estimate, to obtain the QCRB, we only need to focus on the part of the device that couples to those parameters.  In the case of a gravitational-wave detector, we will only need to consider the long arms that contain strong carrier fields, which maps to a simple optomechanical system comprising of a Fabry-Perot cavity with a movable end mirror and containing optical modes. This is the system we consider in this section.

As discussed in Ref.~\cite{pang2018}, in the TT (transverse-traceless) gauge and for a strongly pumped single mode cavity, the gravitational waves interact with the optomechanical cavity according to the Hamiltonian (given here in the interaction picture)
\be\label{HInt}
H_{I}=-\frac{\omega_0\bar{\alpha}}{2}\hat{\alpha}_1(t)\int d^3\B{k}\,J_\lambda(\B{k})\hat{h}_\lambda(t,\B{k})
\ee
with an implicit sum over the GW polarizations $\lambda=+,\,\times$. Here $\omega_0$ is the cavity's resonant frequency, $\hat{\alpha}_1$ is the amplitude quadrature of the cavity mode and $\bar{\alpha}$ is its large classical component. The amplitude quadrature is canonically conjugate to the phase $\hat{\alpha}_2$, and the quadratures $\hat{\alpha}_{1,2}$ are  given in terms of the annihilation and creation operators of the photon mode by 
\begin{equation}
\hat{\alpha}_1=\sqrt{\frac{\hbar}{2}}(\hat{a}+\hat{a}^\dagger)\,,\quad \hat{\alpha}_2=-i\sqrt{\frac{\hbar}{2}}(\hat{a}-\hat{a}^\dagger)
\end{equation}
where $[\hat{a},\hat{a}^\dagger]$=1. In writing down this Hamiltonian we have made the linear and single mode approximations, both of which are valid for strong pumping. We have also decomposed the GW tensor field into its spatial Fourier components such that
\be
\hat{h}_{ij}(t,\B{x})=\int \frac{d^3\B{k}}{\sqrt{(2\pi)^3}}\tau_{ij}^\lambda(\hat{\B{k}})\,\hat{h}_{\lambda}(t,\B{k})e^{i\B{k}\cdot{x}}
\ee
where $\tau_{ij}^\lambda(\bhat{k})$ is the polarization tensor for the $\B{k}$-mode component and depends only on its direction. They are defined by their orthogonality, transverse, and traceless properties :
\be
\tau^\lambda_{ij}\tau^{\lambda'}_{ij}=\delta_{\lambda,\lambda'},\quad
k^i \tau^\lambda_{ij}=0,\quad
\tau^\lambda_{ii}=0
\ee

The term 
\begin{equation}
J_\lambda(\B{k})=\frac{1}{\sqrt{(2\pi)^3}}\tau^\lambda_{xx}(\hat{\B{k}}){\rm sinc}(k_x L/2)e^{ik_xL/2}
\end{equation} represents the antenna pattern which includes the projection of the tensor wave onto the cavity axis (along the $x$ direction), as well as a term accounting for the variation of the GW wave over the length of the cavity. 

The time dependence of the operators reflect the free evolution in the absence of GW interaction under the Hamiltonian terms $H_S$ and $H_B$ for the interferometer and GW field respectively, with
\be
H_S=\frac{\hat{p}^2}{2m}-\frac{\Delta}{2}\left(\hat{\alpha}_1^2+\hat{\alpha}_2^2\right)-\frac{\omega_0\bar{\alpha}}{L}\hat{\alpha}_1\hat{q}+H_{\rm ext}
\ee
and
\be
H_B=\int d^3\B{k}\left[\frac{\big|\hat{\Pi}_\lambda(\B{k})\big|^2}{2M_G}+\frac{1}{2}M_G\omega_k^2\big|\hat{h}_\lambda(\B{k})\big|^2\right]
\ee
Here $H_S$ includes the optomechanical interaction between the cavity mode and movable end mirror ($\hat{p}, \hat{q}$) and $H_{\rm ext}$ accounts for the laser drive. The operator $\hat{\Pi}_\lambda(\B{k})$ is the canonical field momentum for $\hat{h}_\lambda(\B{k})$. The quantity $M_G$ is given by
\begin{equation}
M_G =\frac{c^2}{32\pi G}\,,
\end{equation}
and acts as an effective mass.

For a large amplitude transient the GW field can be decomposed into a classical signal component and quantum fluctuations, so that each $\B{k}$-mode in the signal, which represents a plane gravitational wave, can be written $\hat{h}_\lambda(t,\B{k})=\bar{h}_\lambda(t,\B{k})+\delta \hat{h}_\lambda(t,\B{k})$. Note that waves generated by far away sources are well approximated by plane waves upon their arrival on Earth. All sources of potential interest to LIGO satisfy this condition. Then it is convenient to identify a real parameter representing our signal that is integrated over the wave number and depends only on the propagation direction: 
\be\label{xiPar}
\bar{\xi}^\lambda_{\bhat{k}}(t)=\int \frac{dk}{\sqrt{(2\pi)^3}}\,k^2\bar{h}_\lambda(t,\B{k}){\rm sinc}\left(\frac{kL\cos\theta}{2}\right)e^{ikL\cos\theta/2}
\ee
Here $k=|\B{k}|$, and the analogous quantum fluctuation operator $\hat{\xi}_{\bhat{k}}(t)$ by replacing $\bar{h}_\lambda(t,\B{k})$ in Eq.~\eqref{xiPar} with $\delta\hat{h}_\lambda(t,\B{k})$. We point out $\bar{\xi}_{\bhat{k}}(t)$ is not a pure gravitational wave signal in that it depends on the property $L$ of the probe, which must be therefore be given \textit{a priori}. Suppressing the polarization subscript $\lambda$, the interaction between the probe and the GW field is then
\be\label{HIntAmp}
H_I=-\frac{\omega_0\bar{\alpha}}{2}\hat{\alpha}_1(t)\int d\hat{\B{k}}\;{\tau_{xx}\big(\bhat{k}\big)}\left[\bar{\xi}_{\bhat{k}}(t)+\hat{\xi}_{\bhat{k}}(t)\right]
\ee

\subsection{QCRB for GW transients}
For a particular $\bhat{k}$ direction of the signal and using the linearized Hamiltonian of Eq.~\eqref{HIntAmp}, the signal generator is 
\begin{equation}
\hat{G}_{\bhat{k}}(t)=-{\omega_0\bar{\alpha}\hat{\alpha}_1(t)}\tau_{xx}\big(\bhat{k}\big)/2
\end{equation}
which is proportional to $\hat{\alpha}_1$, and consequently implies that incoming signal will displace the quantum state along $\hat{\alpha}_2$, as is consistent with our understanding of GW detectors' operation. The QFIM is
\be\label{QFIMTime}
F_{\bhat{k}}(t,t')=\frac{4}{\hbar^2}\left(\frac{\omega_0\bar{\alpha}}{2}\right)^2{\left[\tau_{xx}\big(\bhat{k}\big)\right]^2}\,\langle\hat{\alpha}_1(t)\hat{\alpha}_1(t')\rangle_{0,\textrm{sym}}
\ee
where we've assumed that $\langle\hat{\alpha}_1\rangle=0$, as is the case for current GW laser interferometer detectors using a coherent laser drive. In principle Eq.~\eqref{QFIMTime} should contain the covariance of $\hat{\alpha}_1^H$ instead of the interaction picture operator $\hat{\alpha}_1$. However, since $\hat{\alpha}^H_1$ is linear in $\bar{\xi}_{\bhat{k}}$, the signal dependence drops out of $\Delta\hat{\alpha}_1^H$ (ignoring the quantum contribution from $\hat{\xi}_{\bhat{k}}$). For equal time, the inverse of the point QFI $F_{\bhat{k}}(t,t)$ bounds the point estimation error for the signal at time $t$, or
\be\label{pointEstimation}
\langle \Delta\bar{\xi}_{\bhat{k}}(t)^2\rangle\equiv
\Sigma_{\bhat{k}}(t,t)\geq F^{-1}_{\bhat{k}}(t,t)
\ee
To bound the estimation error for the entire waveform, assuming all stationary processes we can diagonalize the QFIM by going to the frequency domain, and write
\be\label{QFIMFreq}
\C{F}_{\bhat{k}}(\Omega)=\frac{4}{\hbar^2}\left(\frac{\omega_0\bar{\alpha}}{2}\right)^2{\left[\tau_{xx}\big(\bhat{k}\big)\right]^2}S_{\alpha_1}(\Omega)
\ee
where $S_{\alpha_1}(\Omega)$ is the symmetrized power spectral density (PSD) for $\hat{\alpha}_1$ and depends only on the input drive. Then the QCRB for GW detectors states that the measurement noise for the $\hat{k}$-mode signal must satisfy the fundamental bound
\be\label{PSDBound}
\int_{-\infty}^{\infty} d\tau\;e^{i\Omega\tau}\Sigma_{\bhat{k}}(t,t-\tau)\equiv
\C{S}_{\bhat{k}}(\Omega)\ge \C{F}^{-1}_{\bhat{k}}(\Omega)
\ee
where $\C{S}_{\bhat{k}}(\Omega)$ is the PSD of the estimation error. We remark that this bound is only concerned with the \textit{quantum} Fisher information and assumes knowledge of $\bhat{k}$, which in practice must be determined through other means (i.e. multiple detectors for sky localization) The detector itself is not sensitive to $\bhat{k}$, and the additional uncertainty with regards to propagation direction can be quantified by the classical Fisher information of its likelihood function.

The bound obtained in Eq.~\eqref{PSDBound} through the QCRB formalism is mathematically equivalent to the  EQL. In fact, the derivation for the EQL~\cite{braginskyQM9} relies on identifying an operator which performs the same function as the SLD.  However, whereas the EQL was physically motivated by a single quantum system measuring a classical force, the QCRB approach offers clarity in situations where the cavity mode interacts coherently with another quantum system such as in WLCs.   It was previously thought that additional fluctuations introduced by the additional quantum systems can limit the devices sensitivity --- yet the QCRB formalism indicates just the opposite: additional fluctuations are in fact {\it necessary} in order to increase the QFIM.  Of course, in order to reach the QCRB, one need to have access to all out-going degrees of freedom. 

Eq.~\eqref{PSDBound} holds whether the interferometer is operating in the tuned or detuned configuration, since the relevant quantity for calculating the QFIM is the interaction term $H_D$, which is the same in either case. Increasing the maximum sensitivity for a GW detector now reduces to tuning three independent quantities: the resonant frequency of the cavity $\omega_0$, the average amplitude of the optical mode $\bar{\alpha}$, and the fluctuations of the amplitude quadrature $S_{\alpha_1}(\Omega)$. We emphasize that this bound is independent of gauge choice (as is necessary for a physical bound), although it is more straightforwardly derived and intuitively understood in the TT gauge, where there is a direct interaction term between the signal and the cavity mode. Notably, the bound is independent of test mass properties to $O(v^2/c^2)$.

\section{Reciprocity between Measurement, Radiation, and Decoherence}\label{sec:Reciprocity}
Let us now more closely investigate the role of the quantum component of the GW field. Allowing the GW field to interact quantum mechanically with the probe leads to two outcomes: i) the dynamical evolution of the gravitational-wave modes due to interaction, and ii) decoherence onto the probe from the field's quantum fluctuations. 

Of course, it is possible to model both without quantization. One can use the equations of semi-classical gravity find classical GW radiation by taking the expectation values for quantum matter, and likewise obtain decoherence from a classical stochastic GW background. However, the results of these classical calculations are different from the quantum treatment. In particular, semiclassical gravity equations predicts zero radiated power through GW waves while the quantum treatment does not. Additionally, a quantum GW bath causes decoherence even in at zero temperature in vacuum. 

\subsection{Radiation of GW Energy}
From Eq.~\eqref{HIntAmp} we obtain the quantum analogue of Einstein's field equations for GW generation. In the regime $r\gg L$, the GWs are approximately radial and their equations of motion are given by
\be\label{GWGen}
\hat{h}_{ij}^{TT}(t,r)=P_{ijxx}\left[\frac{4G}{c^2}\frac{\omega_0\bar{\alpha}}{c^2}\hat{\alpha}_1(t-r/c)\right]
\ee
where $P_{ijxx}$ is the $TT$ projection operator (for a precise definition, see Ref.~\cite{pang2018}). One can show using stress energy conservation that Eq.~\eqref{GWGen} is equivalent to the quadrupole moment formula. Note that in semiclassical gravity one would take the expectation value of the RHS, which would equal zero for a coherent drive. Conversely, the quantum treatment allows us to postpone taking the expectation value until a measurement is made on a physical quantity, which in our case would be the power radiated given by
\begin{equation} 
P_{\rm GW}=-\frac{c^3}{32\pi G}\int d\hat{\B{r}}\, r^2\big \langle \dot{\hat{h}}_{jk}^{TT}\dot{\hat{h}}_{jk}^{TT}\big\rangle\,,
\end{equation}
where $d\hat{\B{r}}$ is the solid angle element and the overdot is the time derivative. Evaluating, we find an expression for the power in terms of the PSD of $\hat{\alpha}_1$ given by 
\begin{equation}
P_{GW}=\frac{32G}{15c^5}\left(\frac{\omega_0\bar{\alpha}}{2}\right)^2\int \frac{d\Omega}{2\pi}\Omega^2S_{\alpha_1}(\Omega).
\end{equation}

 Already in this expression we can recognize there are similarities with $F_{\bhat{k}}(\Omega)$ in Eq.~\eqref{QFIMFreq}, but we can make connection more concrete by recognizing that the number factor of $32G/15c^5$ in $P_{GW}$ includes the effect of TT projection. If we instead write the TT projection explicitly as an integral over the $d^3{\B{k}}$, we find that the power radiated through each channel $\B{k}$ is
\be\label{channelRadiation}
\frac{d}{d^3\mathbf{k}} P_{\rm GW} =\frac{G}{\pi^2c^2}\left(\frac{\omega_0\bar{\alpha}}{2}\right)^2\left[\tau_{xx}(\bhat{k})\right]^2S_{\alpha_1}(\omega_k),\quad
\omega_k=c|\B{k}|
\ee
which can be very succinctly written as
\be
\frac{d}{d^3\B{k}}P_{\rm GW}=\frac{\hbar^2 G}{4\pi^2 c^2}\mathcal{F}_{\bhat{k}} \,.
\ee
This relates power radiated per $d^3\mathbf{k}$ volume to the QFIM.  In terms of the number of quanta radiated, we have
\be
\label{radFisher}
\frac{d}{d^3\B{k}}\dot{\mathcal{N}}=\frac{\hbar}{128\pi^3 M_G\omega_{\mathbf{k}}}\mathcal{F}_{\bhat{k}}  =\frac{1}{8}\frac{(h_{\mathbf{k}}^{\rm zp})^2}{(2\pi)^3}\mathcal{F}_{\bhat k} \,.
\ee

Here we have defined
\begin{equation}
h_{\bf k}^{\rm zp} \equiv \sqrt{\frac{\hbar}{2 M_G \omega_{\bf k}}}
\end{equation}
which represents the level of zero point fluctuation in $h_{\bf k}$.  In this way, the rate at which the device radiates gravitons into the $d^3\mathbf{k}$ space is the inverse of the signal-to-noise ratio measuring $h_{\bf k}$ in terms of its quantum zero point fluctuation. 

\subsection{Decoherence}
Let us now turn to decoherence of the device due to coupling with gravitational waves.

\subsubsection{Master Equation}
To calculate decoherence we begin with the Markovian quantum master equation in the interaction picture 
\begin{equation}
\dot{\rho}_s(t)=-\frac{1}{\hbar^2}\int_0^\infty d\tau\,{\rm Tr}_B\left[H_I(t),[H_I(t-\tau),\rho_s(t)\otimes \rho_B]\right],
\end{equation}
where $\rho_s(t)$ represents the state of the laser interferometer (system) and the trace is performed over the bath state (GW field) given by $\rho_B$. In writing down this equation we are making the usual Born-Markov approximations \cite{gardinerQN}, which assumes that the bath is weakly coupled to the system and that bath's correlation time is very short compared to the interaction picture evolution of the system state as well as the time $t$ of observation. These approximations enable us to write down a differential equation which is local in time. Substituting our interaction Hamiltonian in Eq.~\eqref{HIntAmp} into the general expression, we find
\begin{align}\label{masterFull}
\dot{\rho}_s(t)=&-\left(\frac{\omega_0\bar{\alpha}}{2\hbar}\right)^2\int_0^\infty d\tau\,
\Big\{
\big[\hat{\alpha}_s(t)\hat{\alpha}_1(t-\tau)\rho_s(t)\notag\\
&-\hat{\alpha}_1(t-\tau)\rho_s(t)\hat{\alpha}_1(t)\big] \big\langle\Gamma(t)\Gamma(t-\tau)\big\rangle +h.c
\Big\}
\end{align}
where $\Gamma(t)$ is the bath operator given by
\be\label{bathOperator}
\Gamma(t)\equiv\int d^3\B{k}\left[J(\B{k})\hat{b}(\B{k})e^{-i\omega_kt}+J^*(\B{k})\hat{b}^\dagger(\B{k})e^{i\omega_k t}\right]
\ee
Assuming that the GW field has a white noise spectrum (i.e in a high temperature thermal state) and is therefore $\delta$-function correlated in time such that $\langle \Gamma(t)\Gamma(t-\tau)\rangle=2 \gamma_B\,\delta(\tau)$ where $\gamma_B$ is a constant, the master equation reduces to
\be\label{masterFinal}
\dot{\rho}_s(t)=\left(\frac{\omega_0\bar{\alpha}}{2\hbar}\right)^2\gamma_B\left[2\hat{\alpha}_1\rho_s \hat{\alpha}_1-\{\hat{\alpha}_1^2,\rho_s\}\right]
\ee
We will restrict ourselves to this Markovian case.

\subsubsection{Diffusion in Phase Space}

We can draw some intuition from Eq.~\eqref{masterFinal} by mapping the density matrix to its quasi-probability distribution in phase space using the Wigner transform. Defining the position and momentum operators by 
\begin{equation} 
\hat{a}=(\hat{x}+i\hat{p}/\hbar)\sqrt{2}
\end{equation}
 and its Hermitian conjugate, the Wigner transform is given by
\begin{equation}
{W}(x,p)=1/(2\pi\hbar)\int dy\,e^{ipy/\hbar}\langle x-y/2|\rho_s|x+y/2\rangle\,.
\end{equation}
We point out $\hat{x}$ and $\hat{p}$ do not literally represent spatial position and momentum of the cavity mode, but are in fact proportional to $\hat{\alpha}_1$ and $\hat{\alpha}_2$ modulo factors of $\sqrt{\hbar}$. However, we use the conventional normalization and notation to more easily draw analogies with the well studied massive harmonic oscillator. Then Eq.~\eqref{masterFinal} maps to
\be\label{masterWignerXP}
\frac{\partial}{\partial t}{W}=\hbar\left(\frac{\omega_0\bar{\alpha}}{2}\right)^2\gamma_B\,\frac{\partial^2}{\partial p^2}{W}
\equiv\frac{\hbar^2}{4}D\frac{\partial^2}{\partial p^2}{W}
\ee
The second derivative with respect to $p$ is precisely the decoherence term that destroys correlations between $x$-separated parts of the quantum state \cite{zurekWigner}, for which we've defined the diffusion coefficient
\begin{equation}
D=(4/\hbar)(\omega_0\bar{\alpha})^2\gamma_B.
\end{equation}
The choice of normalization for $D$ will become clear later on.

Let us now consider the bath decay rate $\gamma_B$ more carefully. We can represent it explicitly as a sum of contributions from all angular direction $\bhat{k}$ of the GW field
\begin{align}\label{gammaB}
\gamma_B\,\delta(\tau)=\frac{1}{2} \int d\bhat{k}\int d\bhat{k}'\,
\tau_{xx}(\bhat{k})\tau_{xx}(\bhat{k}')
\Big\langle\hat{\xi}_{\bhat{k}}(t)\hat{\xi}_{\bhat{k}'}(t-\tau)\Big\rangle
\end{align}
We point out that both sides of Eq.~\eqref{gammaB} must be symmetric in $\tau$ and we can therefore replace the unsymmetrized bath correlation on the right-hand side with the symmetrized form. Then, applying the Fourier transform operator $\int_{-\infty}^\infty d\tau\,e^{i\Omega\tau}$ on both sides we find that $\gamma_B$ depends on the cross correlation PSD between the $\hat{\xi}_{\bhat{k}}$ and $\hat{\xi}_{\bhat{k}'}$. Since the bath modes are independent, their PSD must be given by
\begin{equation}
\int_{-\infty}^{\infty} d\tau\;e^{i\Omega\tau}\langle\hat{\xi}_{\bhat{k}}(t)\hat{\xi}_{\bhat{k}'}(t-\tau)\rangle_{\rm sym}=
S_{\bhat{k}}\,\delta(\bhat{k}-\bhat{k}')
\end{equation}
where $S_{\bhat{k}}$ is a constant (note that $S_{\bhat{k}}$ is associated with cross-correlation of $\hat{\xi}_{\bhat{k}}$ operators, while $\C{S}_{\bhat{k}}$ is used to denote the PSD of signal estimation error). Then we have
\be\label{gammaB2}
\gamma_B=\frac{1}{2}\int d\bhat{k}\,\left[\tau_{xx}(\bhat{k})\right]^2 S_{\bhat{k}}
\ee
Substituting Eq.~\eqref{gammaB2} into the decoherence rate $D$ in Eq.~\eqref{masterWignerXP}, we find that $D$ can be resolved into contributions by differential solid angle, or
\be\label{generalDec}
\frac{d}{d\bhat{k}}D=\frac{2}{\hbar}\left(\frac{\omega_0\bar{\alpha}}{2}\right)^2\left[\tau_{xx}(\bhat{k})\right]^2 S_{\bhat{k}}
\ee
Comparing Eq.~\eqref{generalDec} with the expression of QFI given in Eq.~\eqref{QFIMFreq}, we find that $D$ can be directly related to the QFIM of a cavity mode prepared in the vacuum state, for which $\langle \hat{\alpha}^2_1(0)\rangle_0=\hbar/2$. Then we have
\be\label{decVac}
\frac{d}{d\bhat{k}}D={F}^{\rm vac}_{\bhat{k}}(0,0)S_{\bhat{k}}
\ee
where the `${\rm vac}$' superscript indicates vacuum. Eq.~\eqref{decVac} says that the diffusion coefficient governing the probe's evolution is entirely determined by the properties of the bath and the point QFI at initial time. To be clear, Eq.~\eqref{decVac} holds independently of the probe's actual initial state -- the vacuum state's point QFI is simply a quantity that can always be evaluated. Eq.~\eqref{decVac} is therefore a general relation between diffusion and measurement which holds for the interaction Hamiltonian in Eq.~\eqref{HIntAmp} for a white spectrum GW bath with uncorrelated $\bhat{k}$ modes, and is independent of any other system or bath specifics, including their free evolution and initial states.

\subsubsection{Decoherence in position space}

It is interesting to consider whether the QFIM of a particular state can be related to its own decoherence. Let us now consider the particular case where the probe is initially in a superposition of Gaussian wavepackets separated along $x$, for which we will demonstrate that under certain limits its quantum coherence decays linearly at a rate that is determined by the its own QFIM. The initial wavefunction is given by 
\begin{equation}
|\psi_{\rm c}\rangle=\sqrt{N}\left(|\chi_+\rangle+|\chi_-\rangle\right)
\end{equation}
where 
\begin{equation}
\langle x|\chi_{\pm}\rangle=\sqrt{N_{\pm}}\exp\left[-\frac{(x\pm x_0)^2}{2\sigma^2}\right]
\end{equation}
and $N,\,N_{\pm}$ ensures proper normalization. The density matrix has elements $|\chi_{\pm}\rangle\langle\chi_{\pm}|$ and $|\chi_{\pm}\rangle\langle\chi_{\mp}|$ which are respectively the non-coherent and coherent components. The coherent components go to zero in a purely statistical mixture of $|\chi_+\rangle$ and $|\chi_-\rangle$ and represents the presence of quantum superposition. Denoting its Wigner function by $W_{qc}$ (for quantum coherent), we have initially
\be\label{Wc}
W_{\rm qc}(t=0)=\frac{2N}{\pi\hbar}\exp\left[-\frac{x^2}{\sigma^2}-\frac{p^2\sigma^2}{\hbar^2}\right]\cos\left(\frac{2x_0}{\hbar}p\right)
\ee
Substituting Eq.~\eqref{Wc} in the right hand side of Eq.~\eqref{masterWignerXP} and taking the limit $\sigma\rightarrow 0$, we find that the solution to the evolution equation is
\be
{W}_{\rm qc}(t)=W_{\rm qc}(0)e^{-\gamma_{\rm dec}t},\quad
\gamma_{\rm dec}=Dx_0^2
\ee
However, $x_0^2=\langle\Delta\hat{x}^2_s\rangle_{0}=\langle\Delta\hat{\alpha}_{1s}^2\rangle_0/\hbar$, where the $s$ indicates Schroedinger picture operators, and we have
\be\label{gammaDec}
\gamma_{\rm dec}=\frac{2}{\hbar^2}\left(\frac{\omega_0\bar{\alpha}}{2}\right)^2\left[\tau_{xx}(\bhat{k})\right]^2 
\big\langle\left[\Delta\hat{\alpha}_1(0)\right]^2\big\rangle_{0} S_{\bhat{k}}
\ee
Comparing Eqs.~\eqref{gammaDec} and~\eqref{QFIMTime} and denoting the QFIM for the cat state by $F^{\rm c}_{\bhat{k}}$, we have
\be\label{gammaDecQFIM}
\lim_{\sigma\rightarrow0}\frac{d}{d\bhat{k}}\gamma_{\rm dec}
=\frac{1}{2}F^{\rm c}_{\bhat{k}}(0,0)S_{\bhat{k}}
\ee
Thus, the decoherence rate for a superposition of $\hat{x}$-eigenstates is entirely determined by the properties of the GW bath and the state's own point QFI at initial time.

Our results depend on the GW bath having a white spectrum, which is realized when field is in a high temperature thermal state. Generally, $S_{\bhat{k}}$ for a thermal bath at inverse temperature $\beta$ is given by
\be\label{thermalSpectralDensity}
S^{\rm th}_{\bhat{k}}=\frac{1}{(2\pi)^3}\frac{\pi\hbar|\Omega|}{M_Gc^3}\frac{1}{e^{\beta\hbar|\Omega|}-1}\left[{\rm sinc}\left(\frac{|\Omega|\bhat{k}\cdot\B{L}}{2c}\right)\right]^2
\ee
At high temperature $\beta\hbar|\Omega|\ll1$ and to $O(\Omega L/c)^2$, the frequency dependence drops out and we obtain 
$
S^{\rm th}_{\bhat{k}}\approx {4G}/{(\pi\beta c^5)}
$.
Then Eqs.~\eqref{decVac} and~\eqref{gammaDec} take the forms
\be\label{thermalDec}
\frac{d}{d\bhat{k}}D=\frac{4G}{\pi \beta c^5}{F_{\bhat{k}}^{\rm vac}(0,0)},\quad
\frac{d}{d\bhat{k}}\gamma_{\rm dec}=\frac{2G}{\pi \beta c^5}F_{\bhat{k}}^{\rm c}(0,0)
\ee
We observe that the diffusion coefficient and decoherence rate depend only on the inverse bath temperature, the QFI and fundamental constants.

In summary, we have demonstrated that by assuming a white noise spectrum for the GW bath and making the usual Born-Markov approximations for an open quantum system, we obtain an equation of motion for the density matrix of the probe given by Eq.~\eqref{masterFinal}, which has a representation in phase space given by Eq.~\eqref{masterWignerXP}.  From the latter equation we see that while there is no friction term (i.e. no damping to the probe from the GW bath), the probe state undergoes diffusion at a rate $D$ that can be expressed purely in terms of the point QFI of vacuum and the GW bath properties as in Eq.~\eqref{decVac}, which holds for any initial state. For the special case where the probe is initially in a equal superposition of $\hat{\alpha}_1$ eigenstates, this diffusion results in a linear decoherence rate that is determined by the probe's own QFIM, as in Eq.~\eqref{gammaDecQFIM}. The choice of this initial state is motivated by the fact that $\hat{\alpha}_1$ is the Lindblad operator, and it is therefore natural to consider decoherence along the basis of its eigenstates.

\subsection{Discussion of Reciprocity}
\begin{figure}[h]\label{reciprocity}
\includegraphics[scale=0.40]{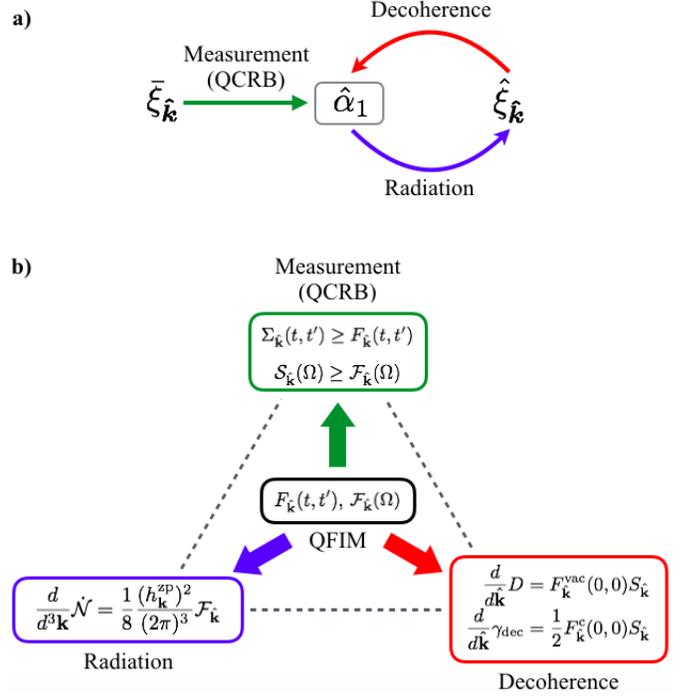}
\centering
\caption{Relating measurement, radiation and decoherence through the QFIM. Figure a) shows the decomposition of the gravitational field into a classical component $\bar{\xi}_{\bhat{k}}$ corresponding to a large amplitude excitation and a quantum component $\hat{\xi}_{\bhat{k}}$. Both components couple to the quantum probe through its degree of freedom $\hat{\alpha}_1$. The probe's interaction with $\bar{\xi}_{\bhat{k}}$ is a measurement for which the fundamental quantum limit on error is given by the QCRB. The probe's interaction with the $\hat{\xi}_{\bhat{k}}$ can be further distinguished into outgoing fluctuations in the form of power radiated as gravitational waves and incoming fluctuations which cause decoherence. Figure b) shows how all three processes can be characterized by the QFIM, given in both the time and frequency domain with respect to a waveform signal (assumed to be stationary). The QCRB which bounds the error of measurement $\Sigma_{\bhat{k}}$ (in time) or $\C{S}_{\bhat{k}}$ (in frequency) is the inverse of the QFIM (Eqs.~\eqref{pointEstimation} and~\eqref{PSDBound}). The power radiated is given by the QFIM and Planck scale constants (Eq.~\eqref{radFisher}). Finally, the diffusion coefficient of the quantum probe in phase space is given by the point QFI of the probe's vacuum  state and the noise spectrum of the bath, and the decoherence rate for a cat state in the eigenbasis of the Lindblad operator $\hat{\alpha}_1$ is the QFI of the cat state itself plus the noise spectrum (Eqs.~\eqref{decVac} and~\eqref{gammaDec}).}
\end{figure}
In this section we have shown for a laser interferometer GW detector that the theoretical limit on its measurement sensitivity, given by its QCRB, is fundamentally related through its QFIM to its radiation of GWs as well as its decoherence from a white noise GW bath (Fig.~\ref{reciprocity} summarizes these relations). It is useful to point out that the power radiation and decoherence correspond to slightly different physical scenarios. 

For radiation, the cavity is being continuously pumped by an external laser drive such that its initial state is forgotten and its amplitude fluctuations has a Fourier transform, which is furthermore assumed to be a stationary process. In this case, the QFIM, expressed in the frequency domain as in Eq.~\eqref{QFIMFreq}, and measured in the level of zero-point fluctuations of the gravitational-wave field, gives the rate at which gravitons are radiated into a unit wavevector space~[Eq.~\eqref{radFisher}]. 
Indeed, the probe must to coupled to an external electromagnetic field in order to vary on timescales relevant for GWs of interest (i.e. at long wavelengths compared to the cavity arm), and should be viewed as a transducer between GWs and the input/output optical fields. From this perspective, Eqs.~\eqref{QFIMFreq} and~\eqref{radFisher} express the idea that the conversion of GWs to photons in signal detection is reciprocal to the conversion of photons to gravitons in radiation. Its channel capacity either as a receiver or a transmitter is fundamentally the same and differ from each other only by a factor involving fundamental constants.

For decoherence, both the GW field and laser drive should be viewed as external baths whose noise fluctuations degrade the purity of the probe's quantum state. However, since the two baths are independent, their effects add in quadrature and we can consider each separately. Therefore, we ignore the effect of laser drive (except to ensure that at time $t=0$ the cavity mode has sufficient photon occupation to linearize the Hamiltonian) and consider only the effect of the GW bath. We find that the probe state diffuses in phase space with coefficient $D$ which is determined by the point QFI of vacuum along with properties of the bath (Eqs.~\eqref{masterWignerXP},~\eqref{decVac}). Interestingly, when the probe is initially in a superposition of $\hat{\alpha}_1$ eigenstates, the decoherence occurs at a linear rate that is determined by the initial cat state's point QFI as in Eq.~\eqref{gammaDecQFIM}. These relations show that the QFI of the probe with respect to a GW signal has a fundamental role in the probe's decoherence due to the quantum fluctuations associated with the signal. 
 
  Interestingly, the processes of radiation and decoherence which result from the quantum component of the field are fundamentally related through the quantum Fisher information (QFI) to the QCRB resulting from field's classical component.  In principle, they can affect our measurement sensitivity. For example, to lower the QCRB for the quantum probe, one must necessarily increase its radiation and decoherence, resulting in an increasingly mixed probe state and a consequent \textit{loss} of sensitivity. Therefore, as we account for the  presence of quantum fluctuations in the signal field,  the QCRB cannot be directly applied to obtain the ultimate maximum sensitivity. How to more precisely formulate the maximum bound in these cases merits further investigation. In the present work, we simply demonstrate the connection for the particular case of GW detection.

\section{Conclusion}
In this work we studied the interactions of a quantum limited laser interferometer, such as LIGO, with a perturbative quantum gravitational-wave field. Using the Hamiltonian derived in \cite{pang2018} and decomposing the gravitational-wave field into its quantum fluctuations and a large excitation interpretable as a classical signal as in Eq.~\eqref{HIntAmp}, we were able to draw fundamental relations between its interactions with the two components of the field. Specifically, we related the three processes of quantum measurement, radiation, and decoherence, with the first process involving the classical component and the latter two involving the the quantum component, which is further divisible into the outgoing and incoming fluctuations corresponding to radiation and decoherence respectively. 

The measurement process is characterized by the quantum Cramer Rao Bound (QCRB) that gives the fundamental quantum limit to measurement sensitivity. The QCRB is equal by the inverse of the detector's quantum Fisher information (or matrix in the multivariate case), and is a property of the quantum probe with respect to the signal it measures. We demonstrated that for our system, this property of the laser interferometer relates the GW power it radiates to fundamental constants (Eq.~\eqref{radFisher}), hence establishing a reciprocal relationship between detection and emission. At the same time, we have shown that under certain conditions the detector's decoherence can be characterized entirely by the QFI and the GW bath properties. Specifically, for a white noise spectrum such as that of a high temperature thermal bath, the detector's diffusion coefficient in phase space is given by the QFI of a cavity at vacuum state (Eq.\eqref{decVac}). Additionally, if the detector is initially prepared in a cat state, it experiences decoherence at a linear rate that is given by its own initial QFI (Eq.~\eqref{gammaDec}). This choice of initial state is highly motivated because it is a superposition of eigenbasis states of the Lindblad operator, i.e. the basis along which decoherence occurs. Although the bath conditions are not completely general, they apply for many systems of interest. 

While these relations have been demonstrated in this work for the specific system of a laser interferometer and gravitational waves, they ultimately derive from the idea that a classical signal has an underlying quantum field which interacts analogously with any quantum measurement device. It is therefore plausible that these relations are generalizable to a broader class of quantum measurement systems, which merits further study.

\acknowledgements 

We thank Bassam Helou, Yiqiu Ma, Haixing Miao, Rana Adhikari, Yuri Levin for discussions.   Our research is supported the  Simons Foundation (Award Number 568762), and the National Science Foundation, through Grants PHY-1708212 and PHY-1708213.

\bibliographystyle{apsrev}
\bibliography{qcrb}
\end{document}